%% file: main.tex
\documentclass[%
 reprint,
preprintnumbers,
nofootinbib,
 amsmath,amssymb,
 aps,
 prl,
]{revtex4-2}

\usepackage{graphicx}
\usepackage{dcolumn}
\usepackage{bm}
\usepackage{hyperref}
\usepackage{xcolor}
\usepackage{natbib}
\usepackage{xspace}
\usepackage{soul}
\usepackage[capitalize]{cleveref}

\usepackage[normalem]{ulem}

\input{abbrev.tex}

\begin{document}

\preprint{DESY-22-018}
\preprint{EFI-22-2}

\title{
New constraints on extended Higgs sectors from the trilinear Higgs coupling}
\author{Henning Bahl$^{1}$}
\email{hbahl@uchicago.edu}
\author{Johannes Braathen$^2$}
\email{johannes.braathen@desy.de}
\author{Georg Weiglein$^{2,3}$}
\email{georg.weiglein@desy.de}
\affiliation{$^1$ University of Chicago, Department of Physics, 5720 South Ellis Avenue, Chicago, IL~60637~USA}
\affiliation{$^2$ Deutsches Elektronen-Synchrotron DESY, Notkestr.~85, 22607 Hamburg, Germany}
\affiliation{$^3$ II. Institut für Theoretische Physik, Universität Hamburg, Luruper Chaussee 149, 22761 Hamburg, Germany}
\date{\today}

\begin{abstract}
    The trilinear Higgs coupling $\lambda_{hhh}$ is crucial for determining the structure of the Higgs potential and for probing possible effects of physics beyond the Standard Model (SM). Focusing on the Two-Higgs-Doublet Model as a concrete example, we identify parameter regions in which $\lambda_{hhh}$ is significantly enhanced with respect to the SM. Taking into account all relevant corrections up to the two-loop level, we show that already current experimental bounds on $\lambda_{hhh}$ rule out significant parts of the parameter space that would otherwise be unconstrained. We illustrate the interpretation of the results on $\lambda_{hhh}$ for a benchmark scenario. Similar results are expected for wide classes of models with extended Higgs sectors.
\end{abstract}

\maketitle


\section{Introduction}

Experimental access to the trilinear Higgs coupling, $\lambda_{hhh}$, is crucial for determining the shape of the Higgs potential and for unravelling the dynamics of the electroweak phase transition. Sizeable deviations from the Standard Model (SM) value are expected in many models of physics beyond the SM (BSM).

Accordingly, one of the main tasks of the Large Hadron Collider (LHC) as well as future colliders is to measure $\lambda_{hhh}$ as precisely as possible, in particular through the process of non-resonant Higgs-boson pair production. Recently, the bound on $\lambda_{hhh}$ has been significantly refined to $-0.4<\kappa_\lambda<6.3$~\cite{ATLAS:2022kbf,ATLAS:2021tyg} at 95\%~C.L., where $\kappa_\lambda\equiv \lambda_{hhh}/\lambda_{hhh}^\text{SM}$, thereby improving the previous best limit~\cite{ATL-PHYS-PUB-2019-009} by a factor of roughly 2. In the future, more precise determinations are expected~\cite{deBlas:2019rxi}: at the high-luminosity LHC, the projected sensitivity for the trilinear Higgs coupling amounts to $0.1<\kappa_\lambda<2.3$ at 95\% C.L.\ with $3\text{ ab}^{-1}$ data~\cite{Cepeda:2019klc} (assuming SM rates); at the ILC and the FCC-$hh$, precision levels of $\mathcal{O}(10\%)$ are expected~\cite{Fujii:2017vwa,Cepeda:2019klc,Goncalves:2018yva}. It should be noted that the sensitivity on $\kappa_\lambda$ can also be affected by BSM contributions to Higgs-boson pair production.

As we will show in this paper, already the current experimental information on $\kappa_\lambda$ puts severe constraints on otherwise unconstrained parameter regions of BSM models with extended Higgs sectors. As a concrete example, we focus on the Two-Higgs-Doublet Model (2HDM). Double-Higgs production in the 2HDM has been studied e.g.\ in Refs.~\cite{Asakawa:2008se,Arhrib:2009hc,Asakawa:2010xj,Baglio:2014nea,Hespel:2014sla,Arhrib:2015hoa,Bian:2016awe,Grober:2017gut,Basler:2017uxn,Basler:2018dac,Babu:2018uik,Basler:2019nas,DiMicco:2019ngk,Alasfar:2019pmn,Arco:2020ucn,Arco:2021bvf,Abouabid:2021yvw}. Moreover, loop corrections to $\lambda_{hhh}$ (often with a focus on non-decoupling effects that can cause large deviations from the SM prediction; the SM result is however recovered in the decoupling limit) have been studied in the 2HDM at the one-loop (NLO)~\cite{Kanemura:2002vm,Kanemura:2004mg,Osland:2008aw,Kanemura:2015mxa,Krause:2016oke,Kanemura:2017wtm} and two-loop (NNLO)~\cite{Braathen:2019pxr,Braathen:2019zoh} levels. Until now, it was, however, believed that deviations of $\lambda_{hhh}$ in the 2HDM were too small to be constrained by existing experimental limits on $\lambda_{hhh}$. We will show that incorporating numerically important two-loop corrections, which we evaluate based on the calculation presented in Refs.~\cite{Braathen:2019pxr,Braathen:2019zoh}, and confronting the obtained predictions with the improved experimental limit of Ref.~\cite{ATLAS:2021tyg} changes this situation. We trace the origin of these large corrections back to large (but still perturbative) trilinear and quartic couplings between the SM-like and the BSM Higgs bosons, which can appear not only in the 2HDM but in many BSM extensions of the SM Higgs sector. As a result for the 2HDM, significant parts of the parameter space that so far were unconstrained are now excluded. 


\section{Constraining the 2HDM parameter space with \texorpdfstring{$\lambda_{hhh}$}{lambda\_hhh}}

We consider a \CP-conserving 2HDM containing two $SU(2)_L$ doublets $\Phi_{1,2}$ of hypercharge $1/2$. We impose a $\mathbf{Z}_2$ symmetry in the Higgs potential under which $\Phi_1\to\Phi_1$, $\Phi_2\to -\Phi_2$, but that is softly broken by an off-diagonal mass term. This potential reads
\begin{align}
\label{eq:HiggsPotential}
& V_{\text{2HDM}}(\Phi_1,\Phi_2) =   \\
&= m_{11}^2\,\Pdd + m_{22}^2\,\Puu - m_{12}^2\left(\Pdu + \Pud\right) \nonumber\\
&\hphantom{=} + \frac{1}{2}\lambda_1 (\Pdd)^2 + \frac{1}{2}\lambda_2 (\Puu)^2  + \lambda_3 (\Pdd)(\Puu) \nonumber        \\
&\hphantom{=} + \lambda_4 (\Pdu)(\Pud) + \frac{1}{2}\lambda_5 \left((\Pdu)^2 + (\Pud)^2\right).\nonumber 
\end{align}
As we focus on the \CP-conserving case, all parameters can be assumed to be real. After minimization of the Higgs potential, the Higgs doublets are decomposed according to $\Phi_i^T = \left(\phi_i^+, (v_i + \phi_i + i \chi_i)/\sqrt{2}\right)$ with $v_1^2 + v_2^2 \equiv v^2 \simeq 246\gev$ and $v_2/v_1 \equiv \tan\beta$. 

Rotating to the mass eigenstate basis, the Higgs boson spectrum consists of the \CP-even Higgs bosons $h$ and $H$ (obtained by a rotation of the $\phi_{1,2}$ states by the angle $\alpha$), the \CP-odd $A$ boson and the neutral Goldstone boson $G$ (obtained by a rotation of the $\chi_{1,2}$ states by the angle $\beta$), as well as the charged Higgs boson $H^\pm$ and the charged Goldstone boson $G^\pm$ (obtained by a rotation of the $\phi_{1,2}^\pm$ states by the angle $\beta$). We identify the lightest \CP-even mass eigenstate $h$ with the observed SM-like Higgs boson and work in the so-called alignment limit by fixing $\alpha = \beta - \pi/2$~\cite{Gunion:2002zf}. This ensures that the tree-level couplings of the $h$ boson are exactly equal to their SM values and in particular that the tree-level trilinear Higgs coupling $\lambda_{hhh}^{(0)}$ is equal to its SM counterpart, $(\lambda_{hhh}^\text{SM})^{(0)} = 3m_h^2/v$. The remaining input parameters for our numerical analysis are $m_H$, $m_A$, $m_{H^\pm}$, $M^2 = m_{12}^2/(\sin\beta\cos\beta)$, and $\tan\beta$. Relations between these parameters and the parameters of \cref{eq:HiggsPotential} are listed e.g.\ in Ref.~\cite{Kanemura:2004mg}.

In order to obtain our predictions we make use of results from Refs.~\cite{Braathen:2019pxr,Braathen:2019zoh,Braathen:2020vwo} for the leading two-loop corrections to $\lambda_{hhh}$ in various BSM models, including an aligned 2HDM. These calculations were performed in the effective-potential approximation, including only the leading contributions involving heavy BSM scalars and the top quark. This implies that we are neglecting all subleading effects from light scalars, light fermions or gauge bosons. Moreover, an on-shell renormalisation scheme is adopted for all the mass parameters that enter the expressions we use, i.e.\ the masses of the top quark and the Higgs bosons, as well as the $\mathbb{Z}_2$ symmetry breaking scale $M$ (for the prescription chosen to determine the counterterm for $M$, we refer to the discussion in Refs.~\cite{Braathen:2019pxr,Braathen:2019zoh}). We find that the largest type of quartic coupling appearing in corrections to $\lambda_{hhh}$ (with one external Higgs boson potentially replaced by the corresponding vacuum expectation value), both at the one- and two-loop level, are those between two SM-like and two heavy BSM Higgs bosons, of the form \begin{align}\label{eq:ghPhiPhi}
    g_{hh\Phi\Phi}=-\frac{2(M^2-m_\Phi^2)}{v^2}\,,
\end{align}
where $\Phi \in {\{ }H, A, H^\pm\}$. We obtain results for $\lambda_{hhh}$ and $\kappa_\lambda=\lambda_{hhh}/(\lambda_{hhh}^\text{SM})^{(0)}$ at the one- and two-loop level.

The limit on $\kappa_\lambda$ obtained in Ref.~\cite{ATLAS:2021tyg} relies not only on the assumption that all other Higgs couplings are SM-like (which is the case in the 2HDM alignment limit) but also that non-resonant Higgs-boson pair production only deviates from the SM via a modified trilinear Higgs coupling. The additional Higgs bosons of the 2HDM can, however, also give rise to further modifications of Higgs-boson pair production. While the resonant contribution with an $H$ ($A$) boson in the $s$ channel is zero in the alignment limit (in the \CP-conserving case) of the 2HDM, at the loop level the additional Higgs bosons can contribute beyond their effects on the trilinear Higgs coupling. However, our calculation includes the leading corrections to Higgs-boson pair production in powers of $g_{hh\Phi\Phi}$ (at NLO and NNLO), which we find to be the source of the large loop corrections in our numerical scan. Therefore, we expect our calculation to capture the dominant effects on Higgs-boson pair production, justifying the application of the experimental limit on $\kappa_\lambda$. 


\section{Numerical results}

While we expect similar results for all 2HDM types,\footnote{The difference between the 2HDM types appears only in the down-type and lepton Yukawa couplings, which play no role in the corrections to $\lambda_{hhh}$ at the level of the leading contributions employed in our calculation.} for our numerical study we concentrate here on the 2HDM of type I. Regarding our predictions for $\kappa_\lambda$, we apply various other constraints of both experimental and theoretical nature on the considered parameter space:
\begin{itemize}
  \item vacuum stability~\cite{Barroso:2013awa} and boundedness-from-below~\cite{Branco:2011iw} of the Higgs potential,
  \item NLO perturbative unitarity~\cite{Cacchio:2016qyh, Grinstein:2015rtl},\footnote{We conservatively demand that $|a_i|<1$ for all eigenvalues $a_i$ of the $2\to2$ scattering matrix.}
  \item electroweak precision observables (EWPO) calculated at the two-loop level using the code \texttt{THDM\_EWPOS}~\cite{Hessenberger:2016atw,Hessenberger:2018xzo},
  \item compatibility of the SM-like scalar with the experimentally discovered Higgs boson using \texttt{HiggsSignals}~\cite{Bechtle:2013xfa,Bechtle:2020uwn},
  \item direct searches for BSM scalars using \texttt{HiggsBounds}~\cite{Bechtle:2008jh,Bechtle:2011sb,Bechtle:2013wla,Bechtle:2020pkv,Bahl:2021yhk},
  \item $b$ physics~\cite{Haller:2018nnx}.\footnote{In practice, the fit results of Ref.~\cite{Haller:2018nnx} are used to obtain $2\sigma$ constraints in the $m_{H^\pm}$--$\tan\beta$ plane of the 2HDM parameter space. }
\end{itemize}
We use \texttt{ScannerS}~\cite{Muhlleitner:2020wwk} to evaluate all of these constraints apart from the NLO perturbative unitarity and the EWPO constraints, which are evaluated separately. If applicable, we demand the constraints to be passed at the $95\%$ C.L. Taking into account these constraints on the parameter space, we obtain for each parameter point the one- and two-loop predictions for $\kappa_\lambda$. We note that as \texttt{ScannerS} does not define a renormalisation scheme for the 2HDM mass parameters, we choose to interpret these as on-shell renormalised inputs when used in the two-loop calculations of the EWPOs and $\lambda_{hhh}$. 


\subsection{Parameter scan}

\begin{figure*}
    \centering
    \includegraphics[width=\textwidth]{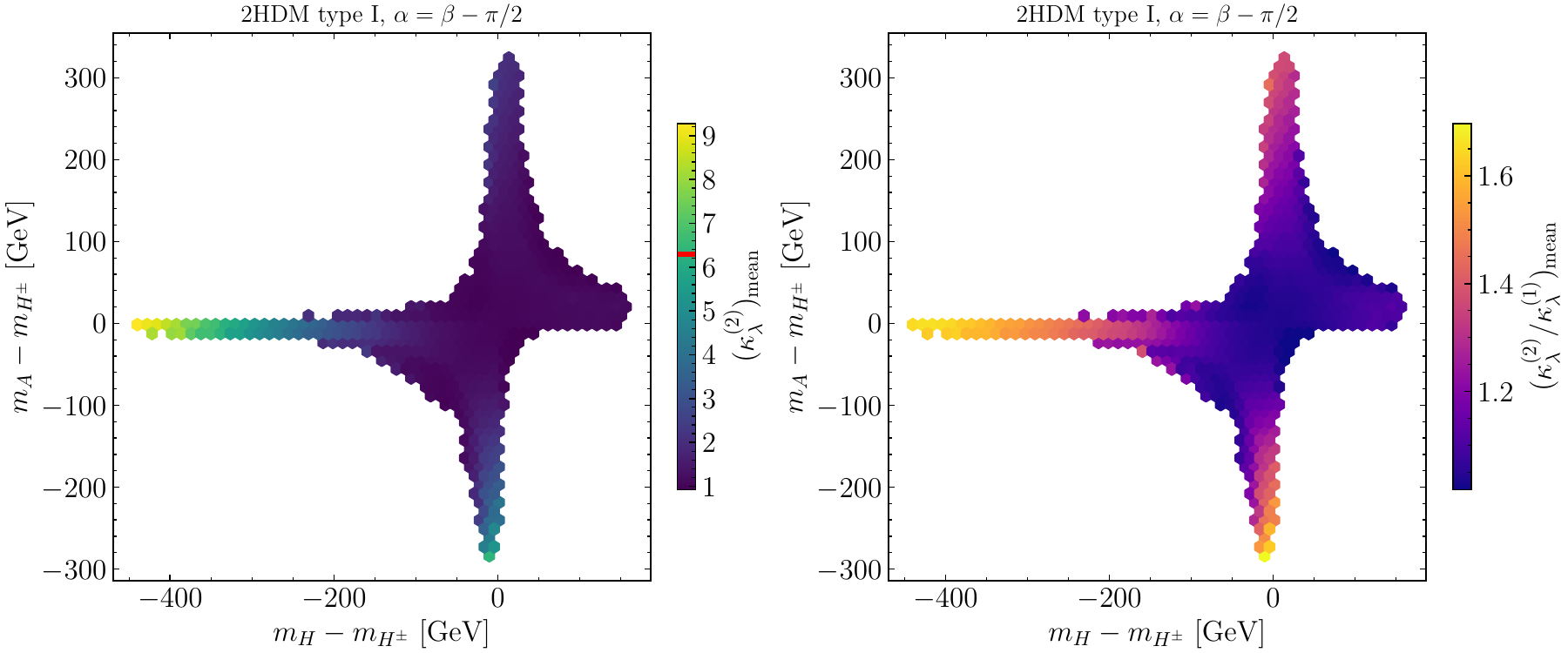}
    \caption{Parameter scan of the type-I 2HDM in the $(m_H - m_{H^\pm}, m_A - m_{H^\pm})$ parameter plane. \textit{Left:} the colour indicates the mean value of $\kappa^{(2)}_\lambda$ in each hexagon-shaped patch; \textit{right:} the colour indicates the mean value of the ratio $\kappa^{(2)}_\lambda/\kappa^{(1)}_\lambda$. In the colour bar of the left-hand plot, the red line indicates the current experimental upper limit on $\kappa_\lambda$.}
    \label{fig:scan}
\end{figure*}

In order to identify the regions with significantly enhanced $\lambda_{hhh}$ we perform a random scan of the 2HDM parameter space. While we fix $m_h=125\text{ GeV}$ and $\alpha=\beta-\pi/2$, we scan over values of the BSM scalar masses in the range $[300\text{ GeV}, 1500\text{ GeV}]$, of $\tan\beta$ between $0.8$ and $50$, and of $m_{12}^2$ between $0$ and $4\cdot 10^6\text{ GeV}^2$. We plot the results of our parameter scan in the $(m_H - m_{H^\pm}, m_A - m_{H^\pm})$ parameter plane in Fig.~\ref{fig:scan}. All shown points pass the theoretical and experimental constraints outlined above. 

In the left panel of Fig.~\ref{fig:scan}, we display for every small hexagon-shaped patch the mean value for $\kappa_\lambda^{(2)}$, which denotes the prediction incorporating contributions up to the two-loop level. This mean value is computed over all the points from the parameter scan contained in each patch. The ``cross-like'' shape of the yet unconstrained region is determined by the electroweak precision constraints which enforce either $m_H \simeq m_{H^\pm}$ or $m_A \simeq m_{H^\pm}$ (see e.g.\ Ref.~\cite{Haber:2010bw}). We find the largest corrections to the trilinear Higgs coupling for $m_A\simeq m_{H^\pm}$ and $m_H-m_{H^\pm}\lesssim -300\text{ GeV}$ and to a lesser extent for $m_H\simeq m_{H^\pm}$ and $m_A-m_{H^\pm}\lesssim -220\text{ GeV}$. In particular, for $m_A\simeq m_{H^\pm}$ and $m_H-m_{H^\pm}\lesssim -375\text{ GeV}$, $\kappa_\lambda$ can reach maximal values of up to $\sim 9.2$. This clearly surpasses the current experimental 95\% C.L. limit of $6.3$, as indicated by the red line in the colour bar of the plot. Accordingly, we find that already the present experimental limits on $\kappa_\lambda$ have an important impact on the viable 2HDM parameter space.\footnote{We note that these mass splittings can also give rise to interesting decay signatures of the BSM Higgs bosons (see e.g.~\cite{Kling:2020hmi,Bahl:2021str,Kling:2022jcd}).} While we have shown in Fig.~\ref{fig:scan} the mean value of $\kappa_\lambda^{(2)}$ for points within each hexagonal patch, it should be noted that showing the minimal value instead would result in a very similar plot. On the other hand, if we consider the maximal values for each patch, we can find large BSM deviations for most of the parameter plane, and in particular also in the centre of the cross shape, where values of $\kappa_\lambda$ above 4 can occur~\cite{Braathen:2019pxr,Braathen:2019zoh}.   

The location of the largest deviations of $\kappa_\lambda$ from the SM can be understood in terms of the interplay between the size of the different underlying couplings entering the corrections to $\lambda_{hhh}$ and the constraints on the allowed 2HDM parameter space. As can be seen from \cref{eq:ghPhiPhi}, the $g_{hh\Phi\Phi}$ couplings grow with the difference between the BSM mass scale $M$ and the masses of the BSM scalars.  On the other hand, while the ``cross-like'' shape of the allowed points is caused by the constraint from EWPO, its boundaries are determined by perturbative unitarity and boundedness-from-below. These two constraints are more stringent in the regions where $m_A < m_H \simeq m_{H^\pm}$ as well as where $m_H>m_A\simeq m_{H^\pm}$ than in the one where $m_H < m_A\simeq m_{H^\pm}$. In terms of model parameters, this translates into smaller allowed splittings between $M$ and the BSM scalar masses, and hence into smaller quartic couplings in the former regions. Consequently, the largest deviations in $\kappa_\lambda$ are then obtained for parameter points where $m_H\simeq M < m_A\simeq m_{H^\pm}$.

After having investigated the absolute size of the corrections to the trilinear Higgs coupling, we assess the relative size of the two-loop corrections in the right panel of Fig.~\ref{fig:scan}. We show there for each hexagon-shaped patch the mean value of $\kappa_\lambda^{(2)}/\kappa_\lambda^{(1)}$ --- the ratio of the two-loop and one-loop predictions for the trilinear Higgs coupling. We find the largest two-loop corrections (in relative size) for $m_H < m_A \simeq m_{H^\pm}$ and to a lesser extent for $m_A < m_H \simeq m_{H^\pm}$ and $m_H \simeq m_{H^\pm} < m_A$. The plot shows that the parameter region where the mean value of $\kappa_\lambda^{(2)}$ is largest coincides with the region where the two-loop corrections are most important, reaching values of close to 70\% of the one-loop corrections. Thus, the proper incorporation of the relevant two-loop corrections is crucial for the confrontation of the prediction for the trilinear Higgs coupling with the experimental bounds. It should be noted that the quite large two-loop corrections encountered here do not indicate a breakdown of perturbation theory. As discussed above, all displayed parameter points pass the criterion of NLO perturbative unitarity. Moreover, employing a dimensional analysis, we have estimated the size of the corresponding dominant three-loop corrections, and find for all points passing all other tests in our scans that the three-loop contributions are estimated to be significantly smaller than the two-loop ones.


\subsection{Benchmark scenario}

\begin{figure}
    \centering
    \includegraphics[width=1.1\columnwidth]{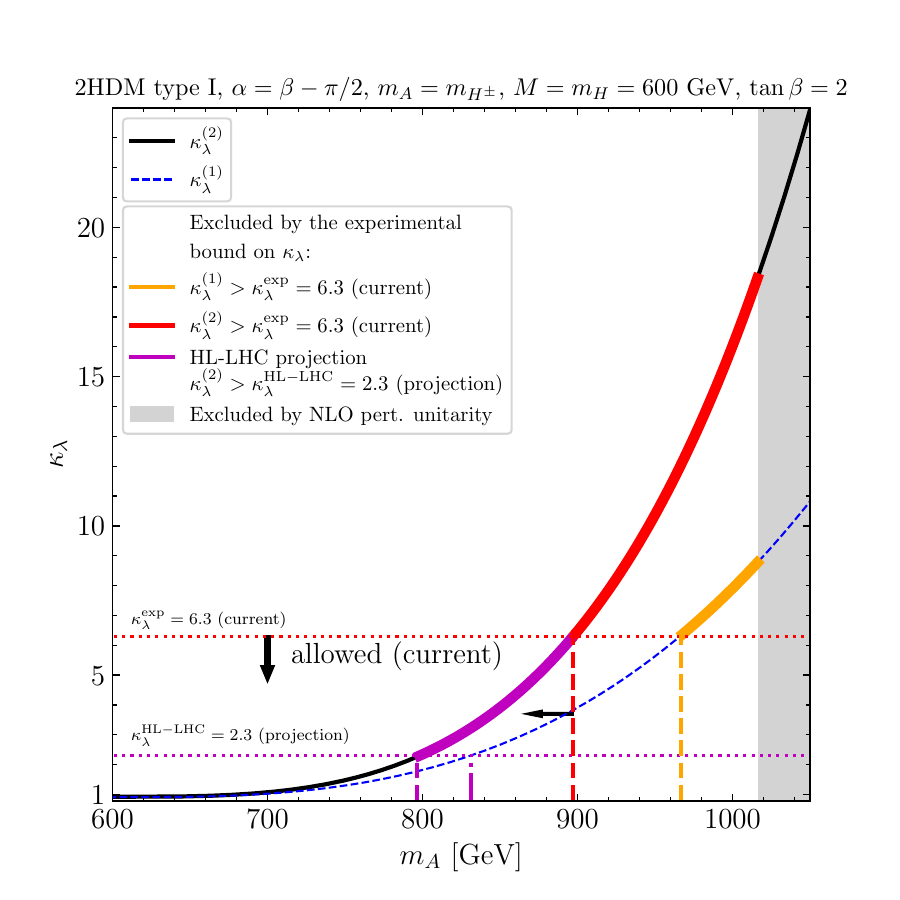}
    \caption{$\kappa_\lambda$ as a function of $m_A$ at one-loop (\textit{dashed blue curve}) and at two-loop order (\textit{solid black curve}). The grey region is excluded by the constraint of NLO perturbative unitarity. The dotted red and purple horizontal lines indicate the current upper limit on $\kappa_\lambda$ and the HL-LHC projection, respectively. The parts of the two- and one-loop curves for $\kappa_\lambda$ that yield a prediction above the current limit of 6.3 are highlighted in red and orange, respectively. The part of the two-loop curve highlighted in purple yields a prediction above the HL-LHC projection for $\kappa_\lambda$.}
    \label{fig:benchmark}
\end{figure}

In order to illustrate the impact of the present (and future) experimental information about $\kappa_\lambda$ on the parameter space of the 2HDM, we consider as an example a benchmark scenario where we fix $M = m_H=600\text{ GeV}$, $m_A = m_{H^\pm}$, $\tan\beta=2$, and $\alpha = \beta - \pi/2$. We then vary $m_A$. We show in Fig.~\ref{fig:benchmark} our results for $\kappa_\lambda^{(1)}$ (dashed blue curve) and $\kappa_\lambda^{(2)}$ (solid black curve) as a function of $m_A$. The colouring indicates parts of the parameter space that are excluded by one or more of the various constraints (or will be probed in the future). In the displayed plot, from the constraints discussed above (besides the one on the trilinear Higgs coupling) only the constraint from NLO perturbative unitarity gives rise to an excluded parameter region, which is displayed in grey. The dotted red and purple horizontal lines indicate the current experimental upper limit on $\kappa_\lambda$ and the projection for the upper limit that the HL-LHC could achieve. The part of the $\kappa_\lambda^{(2)}$ curve highlighted in red indicates the range of masses $m_A$ that is  excluded solely by the current constraint on the trilinear Higgs coupling, where the theoretical prediction includes contributions up to the two-loop level as discussed above. For comparison, we also indicate the part of the $\kappa_\lambda^{(1)}$ curve, highlighted in orange, that would be regarded as excluded if only one-loop contributions were incorporated in the theoretical prediction. Furthermore, the purple-highlighted part of the $\kappa_\lambda^{(2)}$ curve indicates the parameter region that will be probed in the future at the HL-LHC, based on the projection for the upper limit on $\kappa_\lambda$ discussed above.

One can see thatconfronting the existing experimental limit on the trilinear Higgs coupling with state-of-the-art theoretical predictions incorporating contributions up to the two-loop order excludes important parts of the parameter regions of extensions of the SM that would otherwise be allowed by all relevant experimental and theoretical constraints. In the displayed example (with $M = m_H$ kept fixed\footnote{Different choices of $M=m_H$ lead to qualitatively similar results for the same amount of splitting between the masses.} at 600~GeV) the $\kappa_\lambda$ constraint gives rise to an upper limit on $m_A$ of $m_A\lesssim 900$~GeV, while the constraint from NLO perturbative unitarity would allow $m_A$ values of up to 1020~GeV. The impact of the $\kappa_\lambda$ constraint would be much smaller if only the one-loop contributions were included in the theoretical prediction (indicated by the part of the $\kappa_\lambda^{(1)}$ curve that is highlighted in orange). The sensitivity of the HL-LHC in this example will allow one to probe $m_A$ values down to about 800~GeV via an upper limit on $\kappa_\lambda$ or a measurement of a non-SM value. While future data from the LHC will clearly further enhance the impact of the $\kappa_\lambda$ constraint for probing possible scenarios of electroweak symmetry breaking, it should be mentioned that the impact of the theoretical constraint from perturbative unitarity (indicated by the grey area in the plot) is not expected to change in the future. 

Finally, we remark that a more aggressive application of the constraint from
perturbative unitarity would not qualitatively change our results. In particular, demanding that $|\text{Re}(a_i)|<0.5$ for all eigenvalues of the $2\to 2$ scattering matrix would lower the perturbative unitary bound to $m_A \sim  958 \text{ GeV}$, which is still significantly weaker than the current bound imposed by $\kappa_\lambda^{(2)}$. As an additional cross-check, we have verified for several representative points in the benchmark scenario that the scalar couplings do not acquire perturbative-unitarity-violating values under renormalisation-group running until well above the BSM scalar mass scale. We have also confirmed that the inclusion of finite-energy effects in the evaluation of the perturbative unitarity constraint~\cite{Goodsell:2018fex,Goodsell:2018tti} does not lead to more stringent bounds (for these checks, we employed \texttt{SARAH}~\cite{Staub:2009bi,Staub:2010jh,Staub:2012pb,Staub:2013tta} and \texttt{SPheno}~\cite{Porod:2003um,Porod:2011nf}). We leave a detailed study of the perturbative unitarity constraints for future work.


\section{Conclusions}

A precise determination of the trilinear Higgs coupling is crucial for gaining access to the shape of the Higgs potential and for probing possible effects of BSM physics. In this work, we have demonstrated that confronting the latest experimental bounds on the trilinear Higgs coupling with theoretical predictions incorporating numerically important two-loop contributions allows one to exclude significant parts of the parameter space of extensions of the SM Higgs sector that would otherwise seem to be unconstrained. These results have important implications for future searches at the LHC (and elsewhere) and indicate the crucial role played by the trilinear Higgs coupling for discriminating between different possible manifestations of the underlying physics of electroweak symmetry breaking.

Focusing in our numerical discussion on the case of the 2HDM and taking into account other relevant theoretical and experimental constraints, we have found that large BSM quantum corrections can enhance $\lambda_{hhh}$ by up to an order of magnitude as compared to the SM value. We stressed in this context the importance of incorporating a particular class of two-loop corrections, which can reach about 70\% of the one-loop contribution. Based on these findings, we investigated a suitable benchmark scenario and discussed the impact of the present and prospective future bounds on $\lambda_{hhh}$. Our analysis places new exclusion bounds on parameter regions that up to now were in agreement with all relevant constraints.

Further details of our results and their extension to other models with extended Higgs sectors, such as the Inert Doublet Model or a singlet extension of the SM (for which large corrections to $\lambda_{hhh}$ are also known to be possible~\cite{Braathen:2019zoh}), will be presented in an upcoming paper. 


\section*{Acknowledgements}

We thank S.~Hessenberger for providing us access to his code \texttt{THDM\_EWPOS} as well as M. Gabelmann and J.~Wittbrodt  for useful discussions. J.B.\ and G.W. acknowledge support by the Deutsche Forschungsgemeinschaft (DFG, German Research Foundation) under Germany's Excellence Strategy -- EXC 2121 ``Quantum Universe'' – 390833306. H.B.\ acknowledges support by the Alexander von Humboldt foundation.


\bibliography{biblio}

\end{document}

%% file: abbrev.tex
\newcommand{\gev}{\;\text{GeV}\xspace}

\newcommand{\BE}{\begin{equation}}
\newcommand{\EE}{\end{equation}}

\newcommand{\CP}{\ensuremath{\mathcal{CP}}\xspace}

\newcommand{\Pdd}{\Phi_1^\dagger\Phi_1}
\newcommand{\Puu}{\Phi_2^\dagger\Phi_2}
\newcommand{\Pdu}{\Phi_1^\dagger\Phi_2}
\newcommand{\Pud}{\Phi_2^\dagger\Phi_1}
